\documentclass[10pt,aps,prc,twocolumn,superscriptaddress,floatfix]{revtex4-1}
\usepackage{amsmath,amssymb,mathrsfs}
\usepackage{bm}
\usepackage[dvipdfm]{graphicx}
\usepackage{ascmac}
\usepackage{geometry}
\usepackage{color}
\def\bra<#1|{\left<#1\right|}
\def\ket|#1>{\left|#1\right>}
\def\braket<#1>{\left<#1\right>}
\def\bracket<#1|#2>{\left<#1\vphantom{#2}\right|\left.\kern-3pt\vphantom{#1}#2\right>}
\def\braccket<#1|#2|#3>{\left<#1\vphantom{#3}\right|#2\left|\vphantom{#1}#3\right>}

\def\nonum\\{\nonumber\\}

\def\bs#1{\boldsymbol{#1}}
\def\ds#1{\displaystyle{#1}}
\def\bvecr{\boldsymbol{r}}

\def\RE#1#2+#3#4{\mbox{$^{#1}${#2}}+\mbox{$^{#3}${#4}}}
\def\setoftau{s(\{\tau_i\}:V^n\bar{V}^{N-n})}
\def\zI{\mathrm{i}\hspace{0.2mm}}

\begin{document}

\title{
Particle-number projection method in time-dependent Hartree-Fock theory:
Properties of reaction products
}

\author{Kazuyuki Sekizawa}
\email[]{sekizawa@nucl.ph.tsukuba.ac.jp}
%\homepage[]{http://wwwnucl.ph.tsukuba.ac.jp/~sekizawa/english/}
\affiliation{Graduate School of Pure and Applied Sciences, University of Tsukuba, Tsukuba 305-8571, Japan}

\author{Kazuhiro Yabana}
\email[]{yabana@nucl.ph.tsukuba.ac.jp}
%\homepage[]{http://wwwnucl.ph.tsukuba.ac.jp/~yabana/}
\affiliation{Graduate School of Pure and Applied Sciences, University of Tsukuba, Tsukuba 305-8571, Japan}
\affiliation{Center for Computational Sciences, University of Tsukuba, Tsukuba 305-8577, Japan}

\date{\today}

\begin{abstract}
\begin{description}
\item[Background]% This part should describe the context needed to understand what the paper is about.
The time-dependent Hartree-Fock (TDHF) theory has been successful
in describing low-energy heavy ion  collisions. Recently, we have
shown that multinucleon transfer processes can be reasonably
described in the TDHF theory combined with the particle-number
projection technique.
\item[Purpose]% This part should state the purpose of the present paper.
In this work, we propose a theoretical framework to analyze
properties of reaction products in TDHF calculations.
\item[Methods]% This part describes the methods used in the paper.
TDHF calculation in three-dimensional Cartesian grid representation
combined with particle number projection method.
\item[Results]% This part should summarize the results of the methods described in the previous tag.
We develop a theoretical framework to calculate expectation
values of operators in the TDHF wave function after collision
with the particle-number projection. To show how our method
works in practice, the method is applied to $^{24}$O+$^{16}$O
collisions for two quantities, angular momentum and excitation energy.
The analyses revealed following features of the reaction: The nucleon
removal proceeds gently, leaving small values of angular momentum
and excitation energy in nucleon removed nuclei. Contrarily, nuclei receiving
nucleons show expectation values of angular momentum and excitation
energy which increase as the incident energy increases.
\item[Conclusions]% This part should state the conclusions of the paper.
We have developed a formalism to analyze properties of fragment
nuclei in the TDHF theory combined with the particle-number projection
technique. The method will be useful for microscopic investigations
of reaction mechanisms in low-energy heavy ion collisions as well as
for evaluating effects of particle evaporation on multinucleon transfer
cross sections.
\end{description}
\end{abstract}

\pacs{24.10.Cn, 25.60.Je, 25.70.-z, 21.60.Jz}
\keywords{}

\maketitle

\section{INTRODUCTION}

The time-dependent Hartree-Fock (TDHF) theory has been successfully
applied for studying low-energy nuclear reactions: fusion reactions, 
deep inelastic collisions, quasi-fission reactions, extraction of nucleus-nucleus
potentials and dissipation coefficients, and so on (for a recent review,
see Ref.~\cite{Simenel(review)}). For reactions producing projectile-
and target-like fragments, expectation values of operators are often
evaluated in the TDHF wave function after collision to investigate
properties of produced nuclei.

Recently, a particle number projection (PNP) technique has been
proposed by C.~Simenel \cite{Projection}. This method has
made it possible to calculate transfer probabilities efficiently from
the TDHF wave function after collision and has been successfully
applied \cite{Projection2,Scamps,KS_KY_MNT}. Using the PNP
technique, we studied multinucleon transfer (MNT) processes in
heavy ion reactions at around the Coulomb barrier for several
systems \cite{KS_KY_MNT} for which extensive measurements
are available \cite{Corradi(40Ca+124Sn),Corradi(48Ca+124Sn),
Szilner(40Ca+208Pb)2,Corradi(58Ni+208Pb)}. Comparing calculated
cross sections with the measurements, we have concluded that the
TDHF theory may describe MNT cross sections quantitatively,
in an accuracy comparable to calculations by other existing
theories such as GRAZING \cite{GRAZING} and Complex WKB
\cite{CWKB}, which are based on semiclassical approximation,
and a model based on Langevin-type equations of motion
\cite{Zagrebaev(2005),Zagrebaev(2007)1}.

The PNP technique has been utilized to calculate reaction
probabilities which are required to calculate cross sections
of specific nucleon numbers. To investigate reaction mechanisms,
it will be useful to calculate expectation values of operators
in the particle-number projected wave functions. This is
the subject of this article. We will consider a general method
to calculate expectation values of one- and two-body operators.

The method will be also useful to investigate deexcitation
effects on the MNT cross sections. The produced nuclei through
MNT reactions are often highly excited. Measured cross sections
are affected by deexcitation processes such as particle emissions
which take place in relatively longer timescale. In the existing
theories mentioned above, evaporation effects are usually
taken into account employing statistical models. In statistical
models of particle evaporation, excitation energy and angular 
momentum are the basic inputs. The method to be developed
in this paper will be useful to calculate these quantities.

In nuclear structure calculations, it is a routine work to calculate
expectation values of operators in particle-number projected wave
functions \cite{Bender(review)}. In this article, we extend the
formalism to the TDHF wave function after collision with the PNP.
To illustrate how our method works, we analyze properties of
produced nuclei in $^{24}$O+$^{16}$O collisions.

This article is organized as follows. In Sec.~\ref{Sec:formulation},
we describe a general formalism to calculate expectation values of
operators in the TDHF wave function after collision with the PNP.
In Sec.~\ref{Sec:example}, we apply the method to
$^{24}$O+$^{16}$O collisions, as an illustrative example.
In Sec.~\ref{Sec:summary}, a summary is presented.

\section{FORMULATION}{\label{Sec:formulation}}

\subsection{Particle-number projection method}

We consider microscopic TDHF calculations of low-energy heavy
ion collisions in which two fragments, a projectile-like fragment
(PLF) and a target-like fragment (TLF), are produced. In this section,
we develop a general formalism to calculate expectation values of
operators for one of the fragments, either PLF or TLF, with the PNP.
We first describe the formalism assuming that the system is composed
of $N$ identical fermions. An extension to include two kinds of
fermions, neutrons and protons, is straightforward.

We assume that the fragments are well separated spatially after
collision at the final stage of the TDHF calculation. We define
two spatial regions, $V$ and $\bar{V}$. The spatial region $V$
includes a fragment to be analyzed. $\bar{V}$ is the complement
of $V$, which includes the other fragment. 

We denote the TDHF wave function after collision as
$\Psi(x_1,\cdots,x_N)$, where $x$ denotes a set of the spatial
and the spin coordinates, $x \equiv ({\bf r},\sigma)$. 
The wave function $\Psi$ is, in general, not an eigenstate of
the particle-number operator in the spatial region $V$ but
a superposition of states with different particle numbers in $V$.
It can be expressed as
\begin{equation}
\Psi(x_1,\cdots,x_N) = \sum_{n=0}^N \Psi_n(x_1,\cdots,x_N),
\label{Psi_f}
\end{equation}
where $\Psi_n$ denotes a particle-number projected wave function,
\begin{equation}
\Psi_n(x_1,\cdots,x_N) = \hat{P}_n \Psi(x_1,\cdots,x_N).
\end{equation}
$\Psi_n$ is a component of $\Psi$ having $n$ particles in the
spatial region $V$ and $N-n$ particles in the spatial region $\bar{V}$.
The operator $\hat{P}_n$ is the PNP operator defined by
\cite{Projection,KS_KY_MNT}
\begin{eqnarray}
\hat{P}_n
&=&
\hspace{-3mm}\sum_{\setoftau}\hspace{-4mm}
\Theta_{\tau_1}({\bf r}_1)\cdots\Theta_{\tau_N}({\bf r}_N)
\label{projection_op_Theta}\\
&=&
\frac{1}{2\pi} \int_0^{2\pi} d\theta \;e^{\zI (n-\hat{N}_V)\theta},
\label{projection_op}
\end{eqnarray}
where $\setoftau$ indicates that a sum over the sequence
$\tau_1 \tau_2 \cdots \tau_N$ should be taken for all possible
combinations that $V$ appears $n$ times and $\bar{V}$ appears
$N-n$ times. We have introduced a space division function,
$\Theta_\tau({\bf r})$, and a particle-number operator in
the spatial region $\tau$, $\hat{N}_\tau$, which are defined by
\begin{equation}
\Theta_\tau(\bvecr) = \biggl\{
\begin{array}{ccc}
1 & \;\mathrm{if} & \bvecr \in \tau, \\
0 & \;\mathrm{if} & \bvecr \notin \tau,
\end{array}\label{SDF}
\end{equation}
and
\begin{equation}
\hat{N}_\tau = \int_\tau d{\bf r} \sum_{i=1}^N \delta({\bf r}-{\bf r}_i)
= \sum_{i=1}^N \Theta_\tau({\bf r}_i),
\end{equation}
where $\tau$ represents the spatial region either $V$ or $\bar{V}$.

We consider a general operator $\hat{\mathcal{O}}$ and decompose
it into two operators according to the spatial regions:
\begin{equation}
\hat{\mathcal{O}} =
\hat{\mathcal{O}}_V + \hat{\mathcal{O}}_{\bar{V}}.
\end{equation}
The operator $\hat{\mathcal{O}}_V$ represents a part of
the operator $\hat{\mathcal{O}}$ acting to the particle when
it is in the spatial region $V$. The operator $\hat{\mathcal{O}}_{\bar{V}}$
represents the remaining part of the operator $\hat{\mathcal{O}}$.
Any one-body operator which is local in space, $\hat{\mathcal{O}}^{(1)}
=\sum_{i=1}^{N} \hat{o}^{(1)}({\bf r}_i, \sigma_i)$, can be decomposed as
\begin{eqnarray}
\hat{\mathcal{O}}^{(1)}
&=& \sum_{i=1}^{N}
\Bigl( \Theta_V({\bf r}_i) + \Theta_{\bar{V}}({\bf r}_i) \Bigr)\,
\hat{o}^{(1)}({\bf r}_i, \sigma_i) \nonum\\[0.4mm]
&=& \hat{\mathcal{O}}_V^{(1)} + \hat{\mathcal{O}}_{\bar{V}}^{(1)},
\label{op_O_one}
\end{eqnarray}
where $\sigma_i$ denotes the spin coordinate of a particle $i$.
In the same way, a two-body operator, $\hat{\mathcal{O}}^{(2)} =
\sum_{i<j}^N \hat{o}^{(2)}({\bf r}_i, \sigma_i, {\bf r}_j, \sigma_j)$,
can be decomposed as
\begin{eqnarray}
\hat{\mathcal{O}}^{(2)}
&=& \sum_{i<j}^N
\Bigl( \Theta_V({\bf r}_i) + \Theta_{\bar{V}}({\bf r}_i) \Bigr)
\Bigl( \Theta_V({\bf r}_j) + \Theta_{\bar{V}}({\bf r}_j) \Bigr) \nonum\\[-2mm]
&&\hspace{6mm}\times\: \hat{o}^{(2)}({\bf r}_i, \sigma_i, {\bf r}_j, \sigma_j) \nonum\\[0.3mm]
&=& \sum_{i<j}^N \Bigl( 
\Theta_V({\bf r}_i)\Theta_V({\bf r}_j) +
\Theta_{\bar{V}}({\bf r}_i)\Theta_{\bar{V}}({\bf r}_j) \nonum\\[-2.3mm]
&&\hspace{8mm} +
\Theta_V({\bf r}_i)\Theta_{\bar{V}}({\bf r}_j) +
\Theta_{\bar{V}}({\bf r}_i)\Theta_V({\bf r}_j) \Bigr) \nonum\\[0.8mm]
&&\hspace{6mm}\times\: \hat{o}^{(2)}({\bf r}_i, \sigma_i, {\bf r}_j, \sigma_j) \nonum\\[1.5mm]
&=& \hat{\mathcal{O}}_V^{(2)} + \hat{\mathcal{O}}_{\bar{V}}^{(2)}
+ \hat{\mathcal{O}}_{V\bar{V}}^{(2)}.
\label{op_O_two}
\end{eqnarray}
The first (second) term represents two-body interactions
which act when both particles $i$  and $j$ are in the spatial 
region $V$ ($\bar{V}$). The third term represents two-body
interactions which act when a particle $i$ is in the spatial
region $V$ and a particle $j$ is in the spatial region $\bar{V}$.
For wave functions after collision in which two fragments are
well separated, the third term can be ignored if the operator
is short-range two-body interactions. When we calculate
excitation energies of fragment nuclei, we ignore long-ranged
Coulomb interactions acting protons belonging to different
fragments.

The expectation value of the operator $\hat{\mathcal{O}}$ in the
fragment which is composed of $n$ particles and locates in the
spatial region $V$ is given by the expectation value of the operator
$\hat{\mathcal{O}}_V$ in the wave function $\Psi_n$,
\begin{equation}
\mathcal{O}_n^V =
\frac{\bigl<\Psi_n\big|\hat{\mathcal{O}}_V\big|\Psi_n\bigr>}
{\bigl<\Psi_n\big|\Psi_n\bigr>}.
\label{def_On_V_proj}
\end{equation}
The bracket $\bigl<\Psi_n\big|\hat O_V\big|\Psi_n\bigr>$
is defined by
\begin{eqnarray}
&&\hspace{-4mm}
\bigl<\Psi_n\big|\hat{\mathcal{O}}_V\big|\Psi_n\bigr> \nonum\\[1mm]
& \equiv & \int \hspace{-0.5mm} dx_1 \cdots \hspace{-0.5mm} \int \hspace{-0.5mm} dx_N
\,\Psi_n^*(x_1,\cdots,x_N)\,\hat{\mathcal{O}}_V\Psi_n(x_1,\cdots,x_N), \nonum\\[-1mm]
\end{eqnarray}
where the integral over $x$ includes an integration over space and a
sum over spin states, $\int dx \equiv \sum_\sigma \int d\bvecr$. Here
and hereafter, we often use the bracket notation to simplify equations.

The expectation value of the operator $\hat{\mathcal{O}}_V$ without
PNP is given by $\mathcal{O}_V=\bigl<\Psi\big|\hat{\mathcal{O}}_V
\big|\Psi\bigr>$. It is related to $\hat{\mathcal{O}}_n^V$ by
\begin{equation}
\mathcal{O}_V = \sum_{n=0}^N P_n \mathcal{O}_n^V,
\label{O_V_and_On_V}
\end{equation}
where $P_n$ is defined by $P_n = \bigl<\Psi_n\big|\Psi_n\bigr>
=\bigl<\Psi\big|\hat{P}_n\big|\Psi\bigr>$. To derive
Eq.~(\ref{O_V_and_On_V}), we used identities $\sum_{n=0}^N
\hat{P}_n = 1$, $\hat{P}_n \hat{P}_{n'} = \delta_{nn'}\hat{P}_n$,
and $[\hat{\mathcal{O}}_V, \hat{P}_n]=0$.

\subsection{Formulae for the Slater determinant}
{\label{Subsec:expectation_values}}

We present formulae of expectation values which are useful
for the TDHF wave function $\Psi$ given by a single Slater
determinant composed of single-particle wave functions $\psi_i(x)$,
\begin{equation}
\Psi(x_1,\cdots,x_N) = \frac{1}{\sqrt{N!}}
\det\bigl\{ \psi_i(x_j) \bigr\}.
\end{equation}
Using the PNP operator of Eq.~(\ref{projection_op}), the
probability $P_n$ can be calculated as \cite{KS_KY_MNT,Projection}
\begin{eqnarray}
P_n &=& \frac{1}{2\pi} \int_0^{2\pi} 
\hspace{-1.5mm} d\theta \;e^{\zI n\theta}
\bigl<\Psi\big| e^{-\zI\hat{N}_V\theta}\big|\Psi\bigr> \nonum\\
&=& \frac{1}{2\pi}\int_0^{2\pi}\hspace{-1.5mm}d\theta\; e^{\zI n\theta}
\det\mathcal{B(\theta)}.
\label{Pn_projection}
\end{eqnarray}
$\mathcal{B}(\theta)$ denotes a $N$-dimensional matrix,
\begin{equation}
\Bigl( \mathcal{B}(\theta) \Bigr)_{ij} = \int dx\, \psi_i^*(x) \psi_j(x,\theta),
\label{def_B_theta}
\end{equation}
where $\psi_i(x,\theta)$ is defined by
\begin{equation}
\psi_i(x,\theta) \equiv \Bigl( \Theta_{\bar{V}}({\bf r})
+e^{-\zI\theta}\Theta_V({\bf r}) \Bigr) \psi_i(x).
\end{equation}
Using Eqs.~(\ref{projection_op}) and (\ref{def_On_V_proj}),
the expectation value $\mathcal{O}_n^V$ is expressed as
\begin{equation}
\mathcal{O}_{n}^V = \frac{1}{2\pi P_n} \int_0^{2\pi} 
\hspace{-1.5mm} d\theta \;e^{\zI n\theta}
\bigl<\Psi\big|\hat{\mathcal{O}}_V e^{-\zI\hat{N}_V\theta}\big|\Psi\bigr>.
\label{def_On_t}
\end{equation}
In the case of one- and two-body operators, $\hat{\mathcal{O}}_V^{(1)}$
and $\hat{\mathcal{O}}_{V}^{(2)}$, in Eqs.~(\ref{op_O_one})
and (\ref{op_O_two}), expectation values can be calculated by \cite{Shinohara}
\begin{eqnarray}
\mathcal{O}_n^{V(1)} &=&
\frac{1}{2\pi P_n} \int_0^{2\pi} \hspace{-1.5mm} d\theta \;
e^{\zI n\theta} \det \mathcal{B}(\theta) \nonum\\[-1.5mm]
&& \hspace{10mm} \times \;
\sum_{i=1}^N 
\int_V \hspace{-0.8mm} dx\;
\psi_i^*(x) \,\hat{o}^{(1)}(x) \tilde{\psi}_i(x,\theta),
\label{def_On_V_one} \\[1mm]
\mathcal{O}_n^{V(2)} &=&
\frac{1}{2\pi P_n} \int_0^{2\pi} \hspace{-1.5mm} d\theta \;
e^{\zI n\theta} \det \mathcal{B}(\theta)\, \sum_{i<j}^N \nonum\\[-0.2mm]
&& \hspace{-13mm}
\times\,\biggl\{ \int_V \hspace{-0.8mm} dx \int_V \hspace{-0.8mm} dx'
\psi_i^*(x)\psi_j^*(x') \,\hat{o}^{(2)}(x,x')\,
\tilde{\psi}_i(x,\theta)\tilde{\psi}_j(x',\theta) \nonum\\
&& \hspace{-11.5mm}\;
-\int_V \hspace{-0.8mm} dx \int_V \hspace{-0.8mm} dx'
\psi_i^*(x)\psi_j^*(x') \,\hat{o}^{(2)}(x,x')\,
\tilde{\psi}_j(x,\theta)\tilde{\psi}_i(x',\theta) \biggr\}, \nonum\\[1mm]
\label{def_On_V_two}
\end{eqnarray}
where $\tilde\psi_i(x,\theta)$ is defined by
\begin{equation}
\tilde{\psi}_i(x,\theta) \equiv \ds{\sum_{k=1}^N} 
\psi_k(x,\theta) \Bigl( \mathcal{B}^{-1}(\theta) \Bigr)_{ki}.
\label{def_tilde_psi}
\end{equation}
We note that $\{\tilde\psi_i\}$ are biorthonormal to $\{\psi_i\}$,
{\it i.e.} $\int dx\, \psi_i^*(x) \tilde\psi_j(x,\theta) = \delta_{ij}$.

\subsection{Application to the TDHF wave function}

In actual TDHF calculations, the many-body wave function $\Psi$ is
given by a product of two Slater determinants, $\Psi = \Psi_\nu \Psi_\pi$,
where $\Psi_\nu$ is for neutrons and $\Psi_\pi$ is for protons.
We present formulae of expectation values for this wave function.
We denote the PNP operator for neutrons (protons) as
$\hat{P}_N^{(n)}$ ($\hat{P}_Z^{(p)}$), where $N$ ($Z$)
is the number of neutrons (protons) in the spatial region $V$.
The probability that $N$ neutrons and $Z$ protons are in the
spatial region $V$ is then given by a product of probabilities
for neutrons and protons,
\begin{eqnarray}
P_{N,Z} &=&
\bigl<\Psi\big|\hat{P}_N^{(n)}\hat{P}_Z^{(p)}\big|\Psi\bigr> \nonum\\[0.5mm]
&=&
\bigl<\Psi_\nu\big|\hat{P}_N^{(n)}\big|\Psi_\nu\bigr>\,
\bigl<\Psi_\pi\big|\hat{P}_Z^{(p)}\big|\Psi_\pi\bigr> \nonum\\[0.5mm]
&=& P_N^{(n)}P_Z^{(p)}.
\label{def_P_NZ}
\end{eqnarray}

We first consider expectation values for a one-body operator.
We note that any one-body operator can be written as a sum
of operators for neutrons and for protons, $\hat{\mathcal{O}}_V^{(1)}
= \hat{\mathcal{O}}_V^{(1,n)} + \hat{\mathcal{O}}_V^{(1,p)}$.
Thus the expectation value of the one-body operator
$\hat{\mathcal{O}}_V^{(1)}$ is given by a sum of
two terms. For the fragment nucleus specified by $N$ and $Z$,
we have
\begin{eqnarray}
\mathcal{O}_{N,Z}^{V(1)} &=&
\frac{\bigl<\Psi\big|\hat{\mathcal{O}}_V^{(1)}\hat{P}_N^{(n)}\hat{P}_Z^{(p)}\big|\Psi\bigr>}
{\bigl<\Psi\big|\hat{P}_N^{(n)}\hat{P}_Z^{(p)}\big|\Psi\bigr>} \nonum\\[0.7mm]
&=&
\frac{\bigl<\Psi_\nu\big|\hat{\mathcal{O}}_V^{(1,n)}\hat{P}_N^{(n)}\big|\Psi_\nu\bigr>}
{\bigl<\Psi_\nu\big|\hat{P}_N^{(n)}\big|\Psi_\nu\bigr>} +
\frac{\bigl<\Psi_\pi\big|\hat{\mathcal{O}}_V^{(1,p)}\hat{P}_Z^{(p)}\big|\Psi_\pi\bigr>}
{\bigl<\Psi_\pi\big|\hat{P}_Z^{(p)}\big|\Psi_\pi\bigr>}
\nonum\\[1.4mm]
&=&
\mathcal{O}_N^{V(1,n)}+\mathcal{O}_Z^{V(1,p)}.
\label{def_O1_NZ_V_np}
\end{eqnarray}
$\mathcal{O}_{n}^{V(1,q)}$ is defined by
\begin{equation}
\mathcal{O}_{n}^{V(1,q)} = \frac{1}{2\pi P_n^{(q)}} \int_0^{2\pi} 
\hspace{-1.5mm} d\theta \;e^{\zI n\theta}
\bigl<\Psi_q\big|\hat{\mathcal{O}}_V^{(1,q)}
e^{-\zI\hat{N}_V^{(q)}\theta}\big|\Psi_q\bigr>,
\label{def_O1_NZ_V_np_2}
\end{equation}
where $\hat{N}_V^{(q)}$ denotes the particle-number operator for
neutrons ($q=n$) and for protons ($q=p$) in the spatial region $V$.
We will use these formulae, Eqs.~(\ref{def_O1_NZ_V_np}) and
(\ref{def_O1_NZ_V_np_2}), to calculate expectation values of
the kinetic energy operator included in the Hamiltonian and of
the angular momentum operator.

For a two-body operator, expectation values are not simply given
by a sum of neutron and proton contributions, since two-body
operators act between neutrons and protons. Therefore, we apply
the PNP operators for both neutrons and protons simultaneously,
\begin{eqnarray}
\mathcal{O}_{N,Z}^{V(2)}
& = &
\frac{1}{(2\pi)^2 P_N^{(n)} P_Z^{(p)}}
\int_0^{2\pi} \hspace{-1.5mm} d\theta \int_0^{2\pi}
\hspace{-1.5mm} d\varphi \; e^{\zI (N\theta + Z\varphi)}
\nonum\\[0.5mm]
&&\hspace{10mm}\times \,
\bigl<\Psi\bigl|\hat{\mathcal{O}}_V^{(2)}
e^{-\zI (\hat{N}_V^{(n)}\theta+\hat{N}_V^{(p)}\varphi)}\big|\Psi\bigr>.
\label{def_O2_NZ_V_np}
\end{eqnarray}
We will use the above formula to evaluate excitation energy of
nuclei produced through transfer processes.

To evaluate the excitation energy, we need to exclude the energy
associated with the center-of-mass motion. For this purpose,
we  calculate the energy expectation value using
Eqs.~(\ref{def_P_NZ})-(\ref{def_O2_NZ_V_np}) in the
coordinate system which moves with the fragment nucleus.
In practice, we multiply all the single-particle wave functions by
$e^{-\zI{\bf K}_\mu \bs{\cdot}{\bf r}/A_\mu}$, where
${\bf K}_\mu$ is given by ${\bf K}_\mu = M_\mu
\dot{\bf R}_\mu(t_f)/\hbar$, with $M_\mu$, $A_\mu$, and
$\dot{\bf R}_\mu(t_f)$ being the average mass, the average
nucleon number, and the average velocity of the fragment
($\mu =$ PLF or TLF) in the spatial region $V$ at time $t_f$.
We calculate the velocity of the fragment by $\dot{\bf R}_\mu(t_f)
\equiv \bigl[{\bf R}_\mu(t_f+\Delta t)-{\bf R}_\mu
(t_f-\Delta t)\bigr]/(2\Delta t)$.

We denote the calculated energy expectation value in the
fragment nucleus composed of $N$ neutrons and $Z$ protons
as $\mathcal{E}_{N,Z}^V$. We separately achieve ground state
calculations for the fragment nucleus composed of $N$ neutrons
and $Z$ protons, which we denote as $E_{N,Z}^\mathrm{g.s.}$.
We evaluate an excitation energy of the fragment nucleus by
\begin{equation}
E_{N,Z}^{*V}(E,b) \equiv
\mathcal{E}_{N,Z}^V(E,b) - E_{N,Z}^\mathrm{g.s.},
\label{E_ex_nz}
\end{equation}
where $E$ and $b$ denote the incident relative energy and
the impact parameter, respectively.

In the ground state calculation, we employ a mass correction in
the kinetic energy operator, $\frac{\hbar^2}{2m} \rightarrow
\frac{\hbar^2}{2m}(1-\frac{1}{N+Z})$, to take into account
the center-of-mass correction. The same correction is applied in
evaluating the expectation value of the kinetic energy operator using
Eqs.~(\ref{def_O1_NZ_V_np}) and (\ref{def_O1_NZ_V_np_2}),
depending on numbers of neutrons and protons, $N$ and $Z$,
in the fragment nucleus.

\section{AN ILLUSTRATIVE EXAMPLE: $^{24}$O+$^{16}$O~COLLISION}{\label{Sec:example}}

To illustrate usefulness of the PNP method described in
Sec.~\ref{Sec:formulation}, we analyze properties of
fragment nuclei in $^{24}$O+$^{16}$O collisions
described by the TDHF theory. For $^{24}$O, pairing
correlation may be important. In Ref.~\cite{AMEDEE},
the pairing interaction is reported to be negligible in the
ground state, while finite contribution is reported in
Ref.~\cite{TDDM}. In this paper, we restrict ourselves
to treatments ignoring pairing effects. We note that
reactions including neutron-rich oxygen isotopes have
been well-studied in the TDHF theory as a typical reaction
involving light unstable nuclei \cite{Kim(1994),Kim(1997),
UO(3D-FULL),TRF_EPJA}. We will investigate expectation
values of the angular momentum operator and average
excitation energies.

We consider reactions in which two fragments are generated
after collision. We call the $^{24}$O-like fragment nucleus
as the PLF and the $^{16}$O-like fragment nucleus as the TLF.
We describe the collision in the center-of-mass frame. We
choose $xy$-plane as the reaction plane setting the incident direction
parallel to the $x$ axis. The projectile, $^{24}$O, moves towards
the negative-$x$ direction, while the target, $^{16}$O, moves
towards the positive-$x$ direction. The impact parameter vector
is set parallel to the positive-$y$ direction.

\subsection{Computational details}

We use our own computational code of TDHF calculation for
nuclear collisions, as in Ref.~\cite{KS_KY_MNT}. Our code utilizes
a three-dimensional uniform-grid representation for single-particle
wave functions without any symmetry restrictions. The 11-point
high-order finite difference formula is used for the spatial derivatives.
For the time evolution, we use fourth order Taylor expansion
method. The spatial grid points of $N_x \times N_y \times N_z =
90 \times 80 \times 26$ are used with 0.8-fm mesh spacing.

As an initial condition, two nuclei are placed at the distance of
32~fm in the $x$ direction. The initial wave functions of projectile
and target nuclei are prepared in a box with $N_x \times N_y \times N_z
= 40 \times 40 \times 26$ grid points. We calculate time evolution
until a distance between the centers of the PLF and the TLF exceeds
32~fm. For the PNP analysis, integrals over $\theta$ are performed
by employing the trapezoidal rule discretizing the interval $[0,2\pi]$
into $M$ equal grids. We find that $M=30$ is sufficient for the
$^{24}$O+$^{16}$O system. All the results reported here are
calculated using the Skyrme SLyIII.0.8 parameter set \cite{SLyIII}.

\subsection{Ground states}

%&&&&&&&&&&&&&&&&&&&&&&&&&&&&&&&&&&&&&&&&&&&&&&&&&&
\begin{figure}[t]
   \begin{center}
   \includegraphics[width=7.0cm]{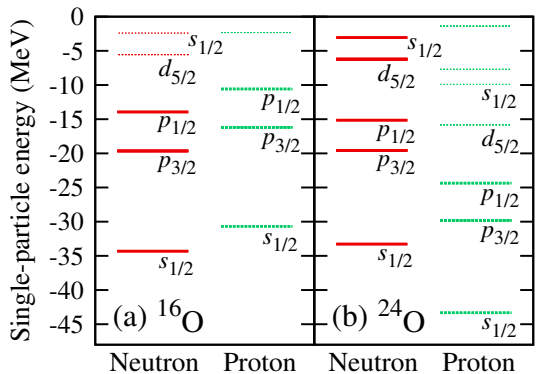}
   \end{center}\vspace{-5mm}
   \caption{(Color online)
   Single-particle energies of occupied orbitals for neutrons (thick red solid
   lines) and protons (thick green dotted lines) in $^{16}$O and $^{24}$O
   are shown in the panels (a) and (b), respectively. Single-particle energies
   of unoccupied orbitals are also shown by thin dotted lines.
   }\vspace{-2mm}
   \label{FIG:spe}
\end{figure}
%&&&&&&&&&&&&&&&&&&&&&&&&&&&&&&&&&&&&&&&&&&&&&&&&&&

We calculate ground states of $^{16}$O and $^{24}$O nuclei,
which are both spherical in the self-consistent solutions.
Figure~\ref{FIG:spe} shows single-particle energies of neutrons
(red solid lines) and protons (green dotted lines) in $^{16}$O
in panel (a) and in $^{24}$O in panel (b). Occupied orbitals are
shown by thick lines, while unoccupied orbitals are shown by
thin lines. As recognized from the figure, there are neutron orbitals
characterized by small binding energies in neutron-rich $^{24}$O
nucleus. All proton orbitals in $^{24}$O are deeply bound.

\subsection{Reaction dynamics}

\begin{figure}[t]
   \begin{center}
   \includegraphics[width=5.5cm]{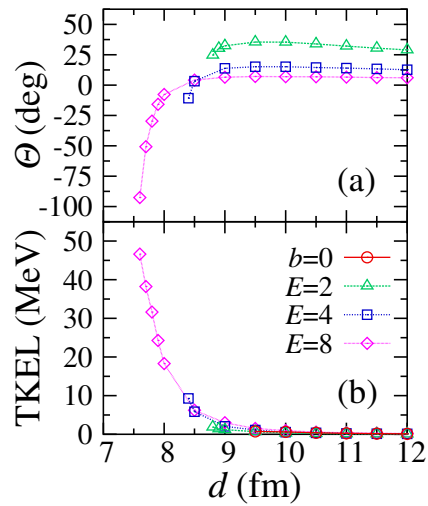}
   \end{center}\vspace{-5mm}
   \caption{(Color online)
   Deflection angle $\Theta$ in the center-of-mass frame (a) and total
   kinetic energy loss (b) are shown as functions of the distance of closest
   approach, $d=d(E,b)$. 
   }\vspace{-2mm}
   \label{FIG:Theta-TKEL_vs_d}
\end{figure}

We first provide an overview of the reaction dynamics in
$^{24}$O+$^{16}$O collisions. In Fig.~\ref{FIG:Theta-TKEL_vs_d},
the deflection angle $\Theta$ in the center-of-mass frame and
the total kinetic energy loss (TKEL) are shown in the panels (a) and (b),
respectively, as functions of the distance of closest approach, $d$.
We evaluate $\Theta$ and TKEL from the momenta of two fragment
nuclei and the Coulomb energy between them at the final stage of
the TDHF calculation where two nuclei are well separated.

We employ the distance of closest approach $d$, instead of
the impact parameter $b$. They are related by
\begin{equation}
d=\frac{Z_\mathrm{P}Z_\mathrm{T}e^2}{2E}+
\sqrt{\Bigl(\frac{Z_\mathrm{P}Z_\mathrm{T}e^2}{2E}\Bigr)^2+b^2},
\end{equation}
where $E$ denotes the incident relative energy. $Z_\mathrm{P}$
and $Z_\mathrm{T}$ denote the proton numbers of the projectile
and the target, respectively. We consider it is useful to use $d$,
because transfer reactions take place at around the distance of
closest approach. For head-on collisions, calculated results
are indicated by $b=0$ and are plotted against $d$ which is related
to the incident relative energy $E$ by $d=Z_\mathrm{P}Z_\mathrm{T}e^2/E$.

We find the fusion reaction takes place at $d=9.4$~fm for head-on
collision ($b=0$) which corresponds to the incident energy of
$E_\mathrm{lab} \sim 24.5$~MeV. For non-central collisions at
incident energies of $E_\mathrm{lab}=2$, 4, and 8~MeV/nucleon,
the fusion reaction is found to take place at $d=8.7$, 8.3, and 7.5~fm,
respectively.

The deflection angle is positive for reactions at the incident energy 
of 2~MeV/nucleon due to the Coulomb repulsion, as seen in
Fig.~\ref{FIG:Theta-TKEL_vs_d}~(a). As the distance of closest
approach decreases, the nuclear attractive interaction acts to
decrease the deflection angle. It becomes negative for $d < 8$~fm
at the incident energy of 8~MeV/nucleon. In the panel (b), we see
an increase of the TKEL at the small-$d$ region where we observed
negative deflection angles.

%&&&&&&&&&&&&&&&&&&&&&&&&&&&&&&&&&&&&&&&&&&&&&&&&&&
%&&&&&&&&&&&&&&&&&&&&&&&&&&&&&&&&&&&&&&&&&&&&&&&&&&

\subsection{Transfer probability}

%&&&&&&&&&&&&&&&&&&&&&&&&&&&&&&&&&&&&&&&&&&&&&&&&&&
\begin{figure}[t]
   \begin{center}
   \includegraphics[width=8.6cm]{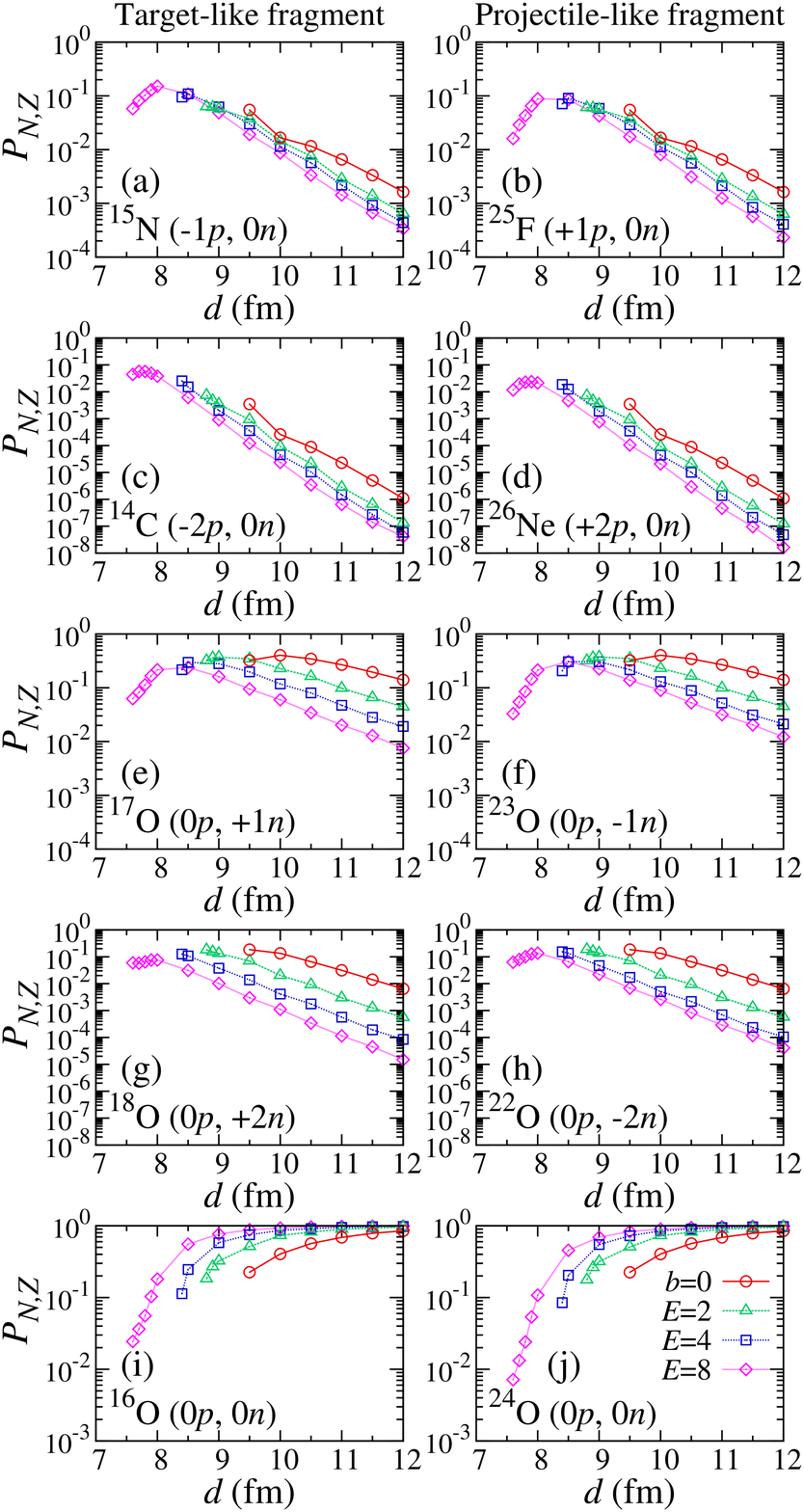}
   \end{center}\vspace{-5mm}
   \caption{(Color online)
   Transfer probabilities with respect to the TLF (left) and the PLF (right)
   are shown as functions of the distance of closest approach, $d=d(E,b)$.
   }\vspace{-2mm}
   \label{FIG:Pnz_vs_d}
\end{figure}
%&&&&&&&&&&&&&&&&&&&&&&&&&&&&&&&&&&&&&&&&&&&&&&&&&&

In Fig~\ref{FIG:Pnz_vs_d}, we show transfer probabilities calculated
using Eq.~(\ref{def_P_NZ}). Red circles show probabilities for head-on
collisions ($b=0$) with several values of $d$. Green triangles, blue
squares, and purple diamonds show probabilities as functions of $d$
for incident energies $E_\mathrm{lab}=2$, 4, and 8~MeV/nucleon,
respectively.

In the calculations, we adopted two choices for the spatial region $V$.
For the probabilities observing a PLF, which are shown in the right
panels of Fig.~\ref{FIG:Pnz_vs_d}, we adopted a sphere with a radius
of 16~fm around the PLF for the spatial region $V$. For the probabilities
observing a TLF shown in the left panels of Fig.~\ref{FIG:Pnz_vs_d},
a sphere with a radius of 16~fm around the TLF is used. We have
confirmed that obtained results are almost independent of the
chosen radius $R$ of the spatial region $V$, if $R$ is taken
in the range of 15~fm $<R<$ 20~fm. We will use this radius
for evaluation of expectation values of angular momentum and
excitation energies.

Figure~\ref{FIG:Pnz_vs_d} (a) and (b) show probabilities of
one-proton transfer processes, while (c) and (d) show probabilities
of two-proton transfer processes. We note that, from the above
choices of $V$ for the PLF and the TLF, the probabilities of proton
removal from $^{16}$O ((a) and (c)) should be coincide with
the probabilities of proton addition to $^{24}$O ((b) and (d)),
if the breakup processes can be neglected. As seen from the
figure, (a) and (b) are very close to each other, indicating that
the breakup processes are indeed negligible. We also find that
(c) and (d) are close to each other. On the other hand, in the
case of neutron transfer channels, one-neutron transfer in panels
(e) and (f) and two-neutron transfer in panels (g) and (h), we find
that the probability of neutron removal from $^{24}$O is much
larger than that of neutron addition to $^{16}$O, especially for
reactions at $E_\mathrm{lab}=$ 8~MeV/nucleon. This fact
indicates that there are substantial probabilities of breakup
processes for neutrons.

In Fig.~\ref{FIG:Pnz_vs_d}, we find that transfer probabilities
decrease as the incident energy increases. Comparing probabilities
of neutron and proton transfer processes, neutron transfer
probabilities are much larger than proton transfer probabilities
at the same distance of closest approach and the same incident
energy. We also find that the slope of probabilities for protons against
the distance of closest approach is much steeper than that for neutrons.
These features are consistent with orbital energies of the two colliding
nuclei in their ground states which are shown in Fig.~\ref{FIG:spe}.
Since there are neutrons bound weakly in $^{24}$O, transfer
probabilities of neutrons are much larger than those of protons.
Since these weekly bound neutrons are spatially extended in
$^{24}$O, we find a long tail of neutron transfer probabilities.

At the highest incident energy of $E_\mathrm{lab}=$ 8~MeV/nucleon,
the proton transfer probability is maximum around $d=8$~fm. The
probability decreases as the distance of closest approach decreases.
The decrease at the small-$d$ region indicates the increase of probabilities
for other channels with transfers of a larger number of protons.

\subsection{Angular momentum}

In this subsection, we investigate expectation values of the
angular momentum operator in the fragment nuclei. We will
use the same definition for $V$ as that in the previous subsection,
spheres with a radius of 16~fm around the center-of-mass of
the PLF and the TLF. We consider the angular momentum
operator in the spatial region $V$, $\hat{\bf J}_V =
\hat{\bf J}_V^{(n)}+\hat{\bf J}_V^{(p)}$. The operator
$\hat{\bf J}_V^{(q)}$ denotes the angular momentum
operator for neutrons ($q=n$) and for protons ($q=p$) in
the spatial region $V$, given by $\hat{\bf J}_V^{(q)} =
\sum_{i \in q} \Theta_V({\bf r}_i)\,\hat{\bf j}_i = \sum_{i \in q}
\Theta_V({\bf r}_i) \bigl[ (\hat{\bf r}_i - {\bf R}_\mu)
\times \hat{\bf p}_i + \hat{\bf s}_i \bigr]$. ${\bf R}_\mu$
is the center-of-mass coordinate of the fragment ($\mu=$~PLF or TLF).

%&&&&&&&&&&&&&&&&&&&&&&&&&&&&&&&&&&&&&&&&&&&&&&&&&&
\begin{figure}[t]
   \begin{center}
   \includegraphics[width=8.6cm]{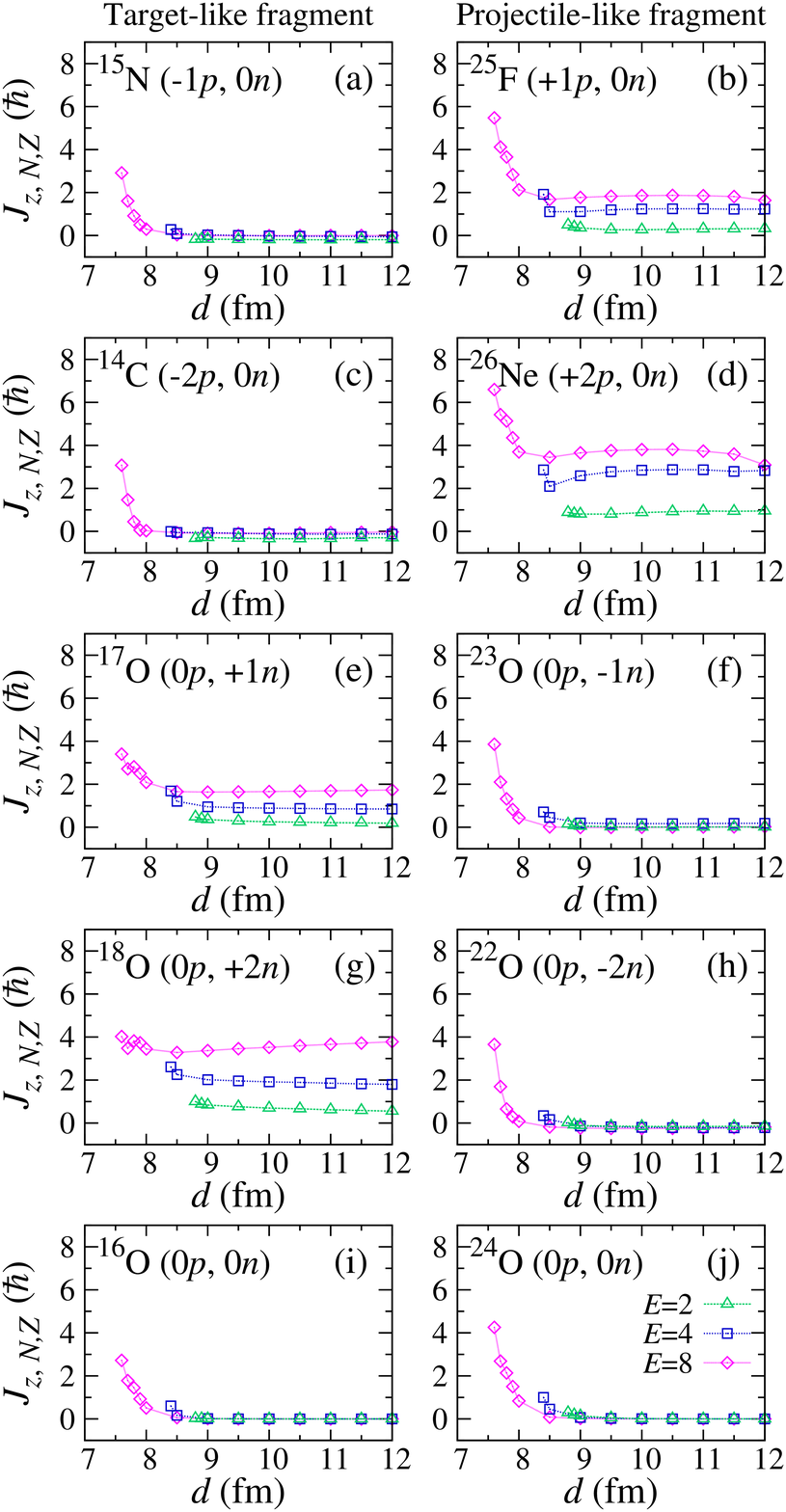}
   \end{center}\vspace{-5mm}
   \caption{(Color online)
   Expectation values of the angular momentum operator for fragment
   nuclei in each transfer channel are shown as functions of the distance
   of closest approach, $d=d(E,b)$.
   }\vspace{-2mm}
   \label{FIG:Jnz_vs_d}
\end{figure}
%&&&&&&&&&&&&&&&&&&&&&&&&&&&&&&&&&&&&&&&&&&&&&&&&&&

Figure~\ref{FIG:Jnz_vs_d} shows expectation values of the angular
momentum operators in the PLF and the TLF composed of specific
numbers of neutrons and protons. A component perpendicular to
the collision plane is shown. Left panels show expectation values in
the TLF, while right panels show those in the PLF. For reactions
at $E_\mathrm{lab}=8$~MeV/nucleon, expectation values at the
small-$d$ region, $d < 8$~fm, are always positive irrespective of the
numbers of transferred nucleons. This fact supports a macroscopic
picture of a friction converting the angular momentum from the
nucleus-nucleus relative motion to the internal ones.

In the following, we discuss results at relatively large-$d$ region
($d > 9$~fm). In these reactions, the distance of closest approach
is much larger than the sum of radii of two colliding nuclei, and
transfer processes are considered to proceed as single-particle
dynamics. TDHF calculations may describe either above-barrier
transfer or quantum tunneling below the barrier. In nucleon
removal channels ((a), (c), (f), and (h)), we find that the expectation
values of the angular momentum operator are very small irrespective
of either neutron(s) or proton(s) is(are) removed, either from
$^{16}$O or $^{24}$O. This fact may be understood from
properties of orbitals. For $^{16}$O, orbitals of the smallest
binding energy are $1p_{1/2}$ for both neutrons and protons.
For $^{24}$O, they are $2s_{1/2}$ for neutrons and $1p_{1/2}$
for protons. We thus find that the orbitals of the smallest binding
energy are characterized by small angular momenta. Since
nucleon removals from spatially extending single-particle orbitals
are expected to take place for orbitals with the smallest binding
energy, removal of nucleons from these orbitals may not leave
large values of angular momentum in nucleon removed nuclei.

In nucleon addition channels ((b), (d), (e), and (g)), we find
finite positive values of angular momentum in all channels.
The expectation values increase as the incident energy increases.
They do not depend much on the distance of closest approach $d$.
These features may be understood by the following intuitive
considerations. Let us consider a transfer of one nucleon from
$^{24}$O to $^{16}$O. We assume that the nucleon transfer
takes place when two nuclei are at the distance of closest approach.
Ignoring the interaction potential by nuclear force, the relative
velocity of two nuclei is approximately given by 
\begin{equation}
v_\mathrm{rel} = \sqrt{ \frac{2}{\mu}
\biggl( E-\frac{Z_\mathrm{P}Z_\mathrm{T}e^2}{d} \biggr) },
\label{v_rel}
\end{equation}
where $E$ and $\mu$ denote the incident relative energy and the
reduced mass, respectively. In the rest frame of $^{16}$O nucleus, 
we assume that the transferred nucleon has the same velocity as
the relative velocity $v_\mathrm{rel}$, ignoring the internal motion
in $^{24}$O. This may be reasonable, since we observed very small
expectation values of the angular momentum in nucleon removed
fragments, as seen in Fig.~\ref{FIG:Jnz_vs_d} (a), (c), (f), and (h).
If the transferred nucleon stays at the surface of $^{16}$O, the
transferred nucleon brings the angular momentum,
\begin{equation}
l_z = R m v_\mathrm{rel},
\label{lz}
\end{equation}
into $^{16}$O, where $m$ is the nucleon mass and $R$ is the
radius of $^{16}$O which we estimate by a simple formula,
$R = r_0 A^{1/3}$, with $r_0 = 1.2$~fm and $A = 16$.

%&&&&&&&&&&&&&&&&&&&&&&&&&&&&&&&&&&&&&&&&&&&&&&&&&&
\begin{figure}[t]
   \begin{center}
   \includegraphics[width=5.5cm]{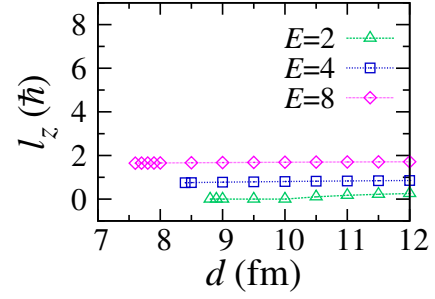}
   \end{center}\vspace{-5mm}
   \caption{(Color online)
   The angular momentum carried into $^{16}$O by an added nucleon
   evaluated by Eqs.~(\ref{v_rel}) and (\ref{lz}) is shown as a function
   of the distance of closest approach, $d=d(E,b)$.
   }\vspace{-2mm}
   \label{FIG:rmvrel_vs_d}
\end{figure}
%&&&&&&&&&&&&&&&&&&&&&&&&&&&&&&&&&&&&&&&&&&&&&&&&&&

In Fig.~\ref{FIG:rmvrel_vs_d}, we show the angular momentum $l_z$
evaluated using Eqs.~(\ref{v_rel}) and (\ref{lz}) as functions of the
distance of closest approach $d$ for several energies. The estimated
values of the angular momentum coincide quantitatively with the
calculated results in channels of one-neutron addition to $^{16}$O,
shown in Fig.~\ref{FIG:Jnz_vs_d}~(e). The estimated angular momentum
depends little on the distance of closest approach $d$, since the Coulomb
potential in Eq.~(\ref{v_rel}) gives only a minor effect except for a case
of very low incident energy. The angular momentum is roughly
proportional to the square root of the energy. In the case of two-nucleon
transfer, the angular momentum carried into $^{16}$O is given by twice
of $l_z$. This reasonably explains the observation in the panel (g).

\subsection{Excitation energy}

%&&&&&&&&&&&&&&&&&&&&&&&&&&&&&&&&&&&&&&&&&&&&&&&&&&
\begin{figure}[t]
   \begin{center}
   \includegraphics[width=8.6cm]{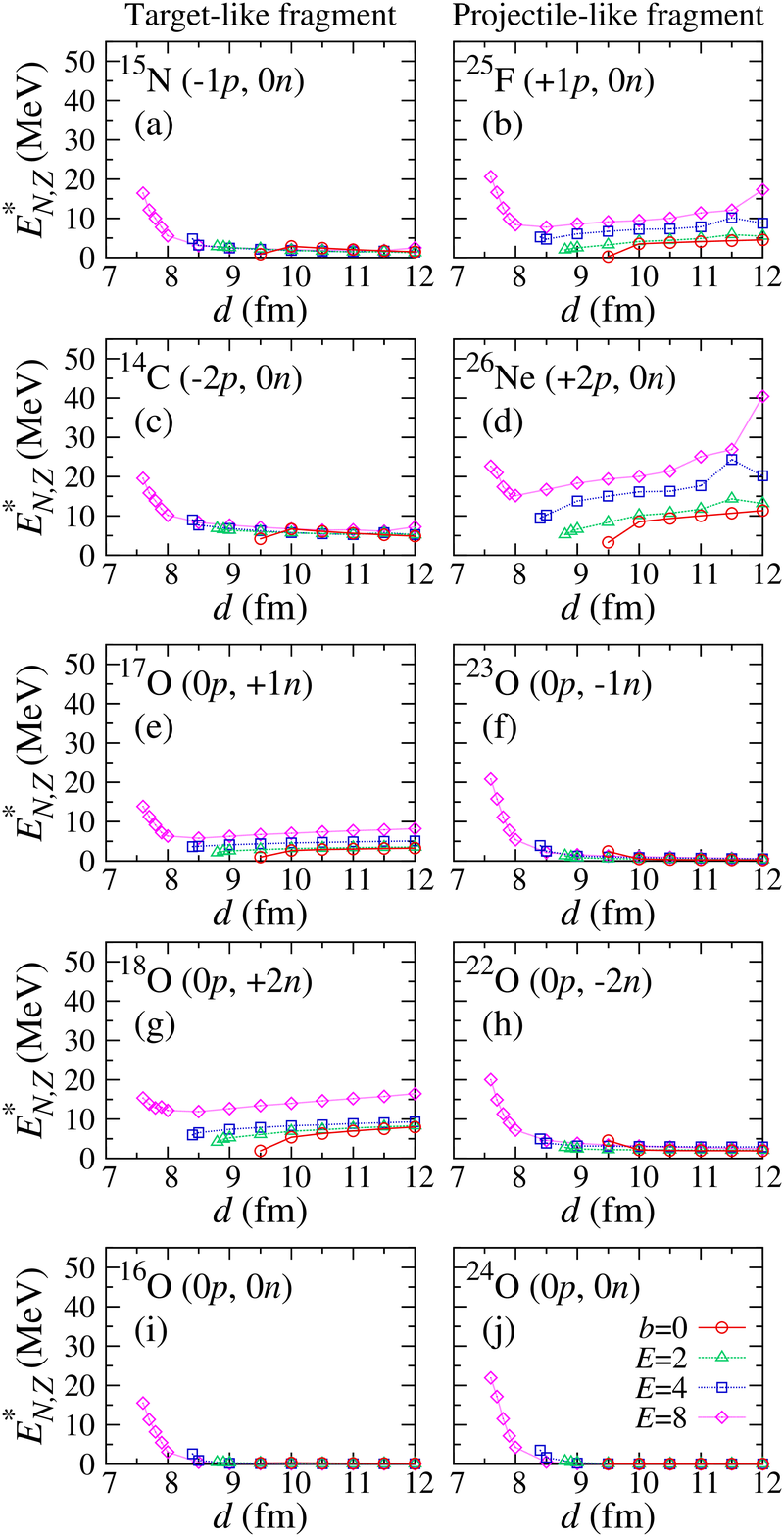}
   \end{center}\vspace{-5mm}
   \caption{(Color online)
   Average excitation energies of fragment nuclei in each transfer channel
   are shown as functions of the distance of closest approach, $d=d(E,b)$.
   }\vspace{-2mm}
   \label{FIG:Eex_vs_d}
\end{figure}
%&&&&&&&&&&&&&&&&&&&&&&&&&&&&&&&&&&&&&&&&&&&&&&&&&&

In Fig.~\ref{FIG:Eex_vs_d}, we show excitation energies of
fragment nuclei evaluated using Eq.~(\ref{E_ex_nz}) as functions
of the distance of closest approach $d$. Left panels show the excitation 
energies of the TLF, while right panels show the excitation energies of
the PLF. As in previous figures, there are two kinds of calculations:
Red circles show results of head-on collisions ($b=0$) varying the
incident energy. Green triangles, blue squares, and purple diamonds
show results for fixed incident energies, $E_\mathrm{lab}=$ 2, 4,
and 8~MeV/nucleon, respectively, changing the impact parameter $b$. 

As we mentioned below Eq.~(\ref{E_ex_nz}), we take into account
the center-of-mass correction in calculating energies of fragment
nuclei and reference energies of ground states in Eq.~(\ref{E_ex_nz}),
while we ignore it in the time evolution calculations. For the quasi-elastic
channels without nucleon transfer, we find very small average excitation
energies at large-$d$ region, $d>9$~fm, as shown in the panels
(i) and (j). This fact may indicate that the inconsistency between
the treatments of the center-of-mass correction in evaluating
excitation energies will not bring any serious problems.

In all channels, we find an increase of the excitation energy in
a small-$d$ region, $d< 8$~fm, where we find an appreciable
TKEL in Fig.~\ref{FIG:Theta-TKEL_vs_d}~(b). At a large-$d$
region, $d > 9$~fm, we have found the small TKEL in
Fig.~\ref{FIG:Theta-TKEL_vs_d}~(b). However, behavior
of the excitation energy depends much on the transfer
channels, as is evident from Fig.~\ref{FIG:Eex_vs_d}.

In nucleon removal channels ((a), (c), (f), and (h)), we find that
excitation energies are rather small. In either one-neutron removal
from $^{24}$O in (f) or one-proton removal from $^{16}$O in (a),
the average excitation energy is less than 3~MeV. This indicates that
the nucleon is removed dominantly from the highest occupied orbital.
In two-nucleon removal channels ((c) and (h)), the excitation energy
becomes somewhat large, about 5-10~MeV in two-proton removal from
$^{16}$O in (c). The excitation energies after nucleon removal are
almost independent of the incident energy. This suggests that nucleons
are removed gently even at higher incident energies.

Contrarily, in nucleon addition channels ((b), (d), (e), and (g)), we
find that excitation energies depend much on the incident energy.
A similar feature was also seen in the angular momentum shown 
in Fig.~\ref{FIG:Jnz_vs_d}, where the added nucleon carries an 
angular momentum associated with the translational relative motion
into the fragment. The expectation values of the angular momentum
were also found to increase as the incident energy increases. This
fact may be related to the increase of the excitation energies as
the incident energy increases in nucleon addition channels: The
transferred nucleons must stay at orbitals of higher angular
momenta as the incident energy increases. The energies of
orbitals with higher angular momenta are high. 

For nucleon addition channels ((b), (d), (e), and (g)), we
observe an increase of excitation energies as the distance of closest
approach increases. One may consider that this fact contradicts
to an intuitive picture that an excitation energy will be smaller as
the distance of closest approach increases since two nuclei cannot
collide violently. We examine this behavior for head-on collisions
($b=0$). As shown by red circles in the panels (b), (d), (e), and (g),
the excitation energies are very small at $d=9.5$~fm. This distance of
closest approach corresponds to slightly outside the boundary of the
fusion reaction. As the distance of closest approach increases (this
corresponds to a decrease of the incident energy in the head-on
collision), the excitation energies increase.

This puzzling behavior can be understood by the following
consideration. As we have shown in Fig.~\ref{FIG:spe}, the Fermi
energies of neutrons and protons in $^{24}$O and $^{16}$O
are rather different because of the excess neutrons in neutron-rich
$^{24}$O. When a nucleon is transferred at a large distance of
closest approach which is much larger than the sum of the radius of 
two colliding nuclei, the nucleon transfer is expected to take place 
between orbitals which are close in energy. The energy-conserving 
transfer processes must cause excitations of produced fragments
if a neutron-rich nuclei is included in the collision.

Let us consider one-proton transfer from $^{16}$O to $^{24}$O
in head-on collisions, which are shown by red circles in the panel (b).
The transfer takes place dominantly for a proton in the highest
occupied orbital of $^{16}$O, $1p_{1/2}$ at $-10.6$~MeV as
shown in Fig.~\ref{FIG:spe}~(a). In Fig.~\ref{FIG:spe}~(b),
we find proton orbitals at a similar orbital energy, $2s_{1/2}$
at $-9.9$~MeV. The proton highest occupied orbital of $^{24}$O
is $1p_{1/2}$ at $-24.3$~MeV and there are $1d_{5/2}$ unoccupied
orbitals at $-15.8$~MeV. Since one of the $1d_{5/2}$ orbitals is
occupied in the ground state of $^{25}$F, we expect the excitation
energy, $E^* \sim \varepsilon(^{24}$O$;\pi 2s_{1/2}) -
\varepsilon(^{24}$O$;\pi 1d_{5/2}) = 5.9$~MeV. This energy
difference almost coincides with the average excitation energy
of $^{25}$F shown in the panel (b) at the large-$d$ region.

We next consider one-neutron transfer from $^{24}$O to $^{16}$O
in head-on collisions, which are shown by red circles in the panel (e).
The highest occupied neutron orbital in $^{24}$O is $2s_{1/2}$ at
$-3.1$~MeV as shown in Fig.~\ref{FIG:spe}~(b). In Fig.~\ref{FIG:spe}~(a),
there are neutron unoccupied orbitals in $^{16}$O at a similar energy,
$2s_{1/2}$ at $-2.4$~MeV. Since the lowest neutron unoccupied orbital
in $^{16}$O is $1d_{5/2}$ orbital at $-5.5$~MeV which is occupied in
the ground state of $^{17}$O, we expect the excitation energy, $E^* \sim
\varepsilon(^{16}$O$;\nu 2s_{1/2}) - \varepsilon(^{16}$O$;\nu 1d_{5/2})
= 3.1$~MeV. This energy difference almost coincides with the average
excitation energy of $^{17}$O shown in  the panel (e) at the large-$d$ region.

In the above considerations, we may understand the transfer mechanism
in terms of orbital properties in the ground state: the highest occupied
orbitals dominantly contribute to the transfer process. We note that,
in Ref.~\cite{TRF_EPJA}, single-particle transfer dynamics in
$^{24}$O+$^{16}$O collision has been examined analyzing
density contributions from individual orbitals. The result reported
in Ref.~\cite{TRF_EPJA} is consistent with the above conclusion.

We make a final comment on an abrupt increase of excitation energy
seen at the largest $d$ value, 12~fm, and the highest incident
energy, 8~MeV/nucleon in panels (b) and (d). We consider that
they are due to a numerical failure. We note that probabilities of
these processes are very small, as confirmed in Fig.~\ref{FIG:Pnz_vs_d}.

\section{SUMMARY}{\label{Sec:summary}}

In the time-dependent Hartree-Fock theory, low-energy
heavy ion collisions are described by a time evolution of
a single Slater-determinant wave function. At the final
stage of calculation, the wave function may be regarded
as a superposition of a number of channels with different
particle numbers and quantum states. To obtain detailed
information on reaction products, projection operator
techniques will be useful. In this paper, we proposed
a method to calculate expectation values of operators
with the particle-number projection to investigate
properties of projectile- and target-like fragments
after collision.

To demonstrate usefulness of our method, we applied the
method to one- and two-nucleon transfer processes in
$^{24}$O+$^{16}$O collisions. We analyzed expectation
values of the angular momentum operator and average
excitation energies of produced nuclei. For fragment
nuclei after nucleon removal, we found small values of
angular momentum and excitation energy, suggesting
a gentle removal of nucleons. For fragment nuclei with
added nucleons, we found substantial expectation values
of angular momentum and average excitation energies.
We have found that the expectation value of the angular
momentum of produced nuclei is proportional to the
relative velocity of the two colliding nuclei at the turning
point. The excitation energy can be understood by a
transfer of nucleons between approximately degenerate
orbitals of projectile and target nuclei.

The above example clearly shows the usefulness of the present
method for microscopic investigations of reaction mechanisms
in heavy ion collisions. The formalism will also be useful to
estimate effects of particle evaporation after multinucleon
transfer processes, which are difficult to describe directly in
the time-dependent Hartree-Fock calculation because of the
very long timescale of the evaporation processes \cite{KS_KY_FUSION14}.

\begin{acknowledgments}
K.S. would like to thank K.~Washiyama and S.A.~Sato for
valuable comments and discussions. K.S. greatly appreciates the
organization of the TALENT course \#6 ``Theory for exploring
nuclear reaction experiments" held at GANIL, Caen, France, in
July 2013, which stimulates a part of analyses in the present work.
This research used computational resources of the HPCI system
provided by Information Initiative Center, Hokkaido University,
through the HPCI System Research Project (Project ID: hp140010).
This work was supported by the Japan Society of the Promotion of
Science (JSPS) Grants-in-Aid for Scientific Research Grant Numbers
23340113 and 25104702, and by the JSPS Grant-in-Aid for JSPS
Fellows Grant Number 25-241.
\end{acknowledgments}

\end{document}